\documentclass{article}
\usepackage[letterpaper]{geometry}
\usepackage{hyperref}
\usepackage{mathtools}
\usepackage{multirow}
\usepackage{algorithm,algpseudocode}
\usepackage{stfloats}

\newcommand{\algoname}{{\sc CusTaRd}}
\newcommand{\dampen}{redirection}
\begin{document}

\title{Random Walks with Variable Restarts\\for Negative-Example-Informed Label Propagation}
\author{Sean Maxwell \\ {\small Department of Computer and Data Sciences} \\ {\small Case Western Reserve University} \\ \href{mailto:stm@case.edu}{stm@case.edu} \and Mehmet Koyut\"urk \\ {\small Department of Computer and Data Sciences} \\ {\small Case Western Reserve University} \\ \href{mailto:mxk331@case.edu}{mxk331@case.edu}}

\date{}
\maketitle

\begin{abstract}
Label propagation is frequently encountered in machine learning and data mining 
applications on graphs, either as a standalone problem or as part of node classification.
Many label propagation algorithms utilize random walks (or network propagation), which provide limited
ability to take into account negatively-labeled nodes (i.e., nodes that are known to be not associated with the label of interest). 
Specialized algorithms to incorporate negatively labeled samples generally focus on learning or readjusting the edge weights to drive walks away from negatively-labeled nodes and toward positively-labeled nodes.
This approach has several disadvantages, as it increases the number of parameters to be learned, 
and does not necessarily drive the walk away from regions of the network that are rich in negatively-labeled nodes.  

We reformulate random walk with restarts and network propagation to enable ``variable restarts", that is the increased likelihood
of restarting at a positively-labeled node when a negatively-labeled node is encountered.
Based on this reformulation, we develop \algoname, an algorithm that effectively combines variable restart probabilities and edge re-weighting to avoid negatively-labeled nodes.
In addition to allowing variable restarts, \algoname\ samples negatively-labeled nodes from neighbors of
positively-labeled nodes to better characterize the difference between positively and negatively labeled nodes.
To assess the performance of \algoname, we perform comprehensive experiments on four network datasets
commonly used in benchmarking label propagation and node classification algorithms.
Our results show that  \algoname\ consistently outperforms competing algorithms that learn/readjust edge weights, 
and sampling of negatives from the close neighborhood of positives further improves predictive accuracy.


\end{abstract}

\section{Introduction}
Label propagation is a commonly encountered problem in data mining and machine learning applications on network and graph-structured data~\cite{garza2019community,pmlr-v80-wagner18a}.
The problem entails assigning labels to nodes of a graph based on knowledge of the labels of a set of ``seed" nodes, such that nodes that are proximate to seed nodes are assigned similar labels.
Label propagation can be considered a special case of the node classification problem, 
in which only graph topology is used in predicting the labels of the nodes.
In contrast, in the general setting for node classification, additional features are available~\cite{wang2021semi}.

\vspace{0.02in}
\noindent
{\bf Label propagation and machine learning on graphs:}
While many machine learning algorithms have-been developed for semi-supervised node classification in the last few years, label propagation is often encountered as part of node 
classification~\cite{APPNP}.
In many cases, the set of training samples can be too small for effective learning, 
thus label propagation is applied prior to training more sophisticated learning 
algorithms~\cite{li2018deeper}.
In addition, emerging evidence suggests that combination of label propagation with simple
models often outperforms more sophisticated models, such as graph neural 
networks~\cite{huang2021combining}.
Despite the ubiquity of label propagation in supervised learning, 
efforts on effectively utilizing negatively-labeled examples in label propagation 
have been relatively scarce.

\vspace{0.02in}
\noindent
{\bf Existing approaches to negative-example-informed label propagation:}
Many label propagation algorithms utilize random walks and their variants~\cite{FU201485, hwang2010heterogeneous, zhou2004, xie2021rwsf, lin2016}.
While classical random walks work with only positively-labeled examples, it has been shown that the utilization of negatively-labeled examples in training improves the accuracy of label propagation~\cite{PNLP}.
Existing approaches to informing random walks with negative examples  use optimization to learn edge weights~\cite{SRW,QUINT} or restart probabilities~\cite{RWER,TELETUNE} that minimize flow into 
negatively-labeled nodes.
Since the number of edges in a network is much larger than the number of nodes,
the number of parameters that need to be learned is usually very large, making learning-based
approaches vulnerable to over-fitting.
In addition, the optimization problems are often non-convex and prone to getting stuck at local optima.

\vspace{0.02in}
\noindent
{\bf Our contributions:}
We improve negative-example-informed label propagation in two ways.
Firstly, we propose a new method that {\em combines re-weighting of edges with variable restart probabilities during label propagation}.
For this purpose, we reformulate random walks to model restarts as part of the network topology, i.e., as directed edges from any node to the positively-labeled nodes.
We then use this formulation to readjust edge weights such that the flow into negatively-labeled nodes is redirected as restarts to positively-labeled nodes. 
The resulting algorithm, \algoname, utilizes negatively-labeled nodes within the random-walk/network-propagation framework and a parameter controlling the aggressiveness of redirection to
reduce the flow into negatively-labeled nodes, 
without requiring training or optimization of a large number of parameters.

Secondly, we propose a {\em positive-neighborhood based approach to sampling negative examples}.
This approach is motivated by the observation that negatively-labeled samples may not always be available or the number of negatively labeled examples is often much larger than positively-labeled examples, thus sampling of a smaller set of negative examples is usually needed.
As opposed to sampling uniformly from the entire set of negative examples, we propose
sampling negative examples from the close neighborhood of positive-examples.
This approach is motivated by the notion that the algorithm can better learn how to distinguish
positives from negatives if it is presented by negatives that are {\em similar} to the positives.
In our experimental studies, we comprehensively investigate the merit of this approach in the context of the proposed algorithm, as well as competing algorithms.

\vspace{0.02in}
\noindent
{\bf Organization of the paper:}
In the next section, we define the label propagation problem and describe random walk and
network propagation based algorithms for label propagation.
Subsequently, we reformulate random walks to enable variable restarts, and show how this reformulation allows readjustment of edge weights into restart probabilities. 
We then describe our approach to sampling negatively-labeled nodes. 
In Section~\ref{sec:results}, we start by describing the four datasets we use for validation,
competing algorithms, and our experimental setup.
We then present the comparison of the predictive accuracy of \algoname\ and competing algorithms
as well as their robustness to scarcity of training examples,
characterize the effect of the \dampen~factor on \algoname's performance, and comprehensively
investigate the effect of the sampling of negative examples on the performance of all algorithms considered.
We conclude our discussion and outline future avenues for research in Section~\ref{sec:discussion}.

\section{Methods}

\subsection{Problem Definition and Existing Approaches.}

Let $G = (V, E)$ denote a graph/network with node set $V$ and edge set $E$.
The nodes in $V$ are associated with categorical ``labels'', where the nodes in subset $S_i \subset V$ are associated with label $i$.
There may be multiple labels available, and $\mathcal{S} = \{S_1,S_2,...,S_k\}$ denotes the set of all available label sets. 
This information is usually incomplete, i.e., $\cup_{i=1}^{n}S_i \neq V$. 
A common problem is ``label prediction'' which, given the labels in $\mathcal{S}$, 
is the task of predicting labels for the unlabeled $v \in V$. 
This problem is often approached using label propagation.

\begin{figure}[t]
    \centering
    \includegraphics[width=0.5\textwidth]{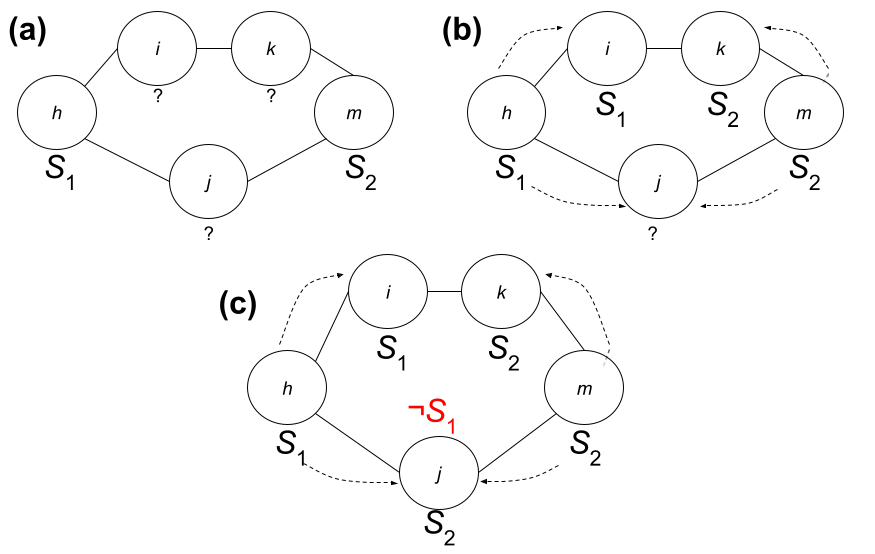}
    \caption{{\small{{\bf The formulation of the label propagation problem.}
    (a) The general setting for label propagation where nodes can be labeled using multiple labels (shown under the nodes) and we want to predict labels for unlabeled nodes (b) The known labels are propagated from each node and the most likely label is assigned to unlabeled nodes. In this case it is difficult to predict a label for node $j$ because it is equally proximal to nodes $h$ and $m$. (c) Label propagation with negatively-labeled examples. The negative label for node $j$ (shown above the node) informs the prediction that node $j$ should not be labeled by $S_1$ so it is labeled by $S_2$.}}}
    \label{fig:labelprop}
\end{figure}

In label propagation, nodes $v \in S_i$ share their label information with their neighbors, who in turn share with their neighbors etc. to ``propagate'' node labels across the network~\cite{raghavan2007, zhou2004}.
The algorithms used to propagate labels are similar to the algorithms used for network propagation, where rather than discrete valued labels, network propagation focuses on propagating continuous values such as flow or probability across a network~\cite{cowen2017}.
Random walk with restarts is a commonly utilized network propagation method that simulates a random walk
across the network by making frequent restarts at the nodes labeled by $S_i$.

\vspace{0.02in}
\noindent
{\bf Random walk with restarts (RWR):}
To formulate RWR, let $\mathbf{A}$ denote the adjacency matrix of $G$.
We use $\mathbf{A}_{i,j}$ to denote matrix entries, $\mathbf{A}_{i,:}$ for rows and $\mathbf{A}_{:,j}$ for columns.
Given $S_i$, were refer to the nodes $v \in S_i$ as {\em seed nodes}.
RWR~\cite{RWR} propagates the labels of $S_i$ to other nodes of $G$ using a column stochastic transition matrix $\mathbf{A}^{(cs)}$ derived from $\mathbf{A}$ defined as $\mathbf{A}^{(cs)}_{i,j}=\mathbf{A}_{i,j}/\sum_{k}\mathbf{A}_{k,j}$.
A restart vector $\mathbf{r}_i$ is used to localize the random walk around the seed nodes, where
$\mathbf{r}_i(v) = 1/|S_i|$ for $v \in S_i$ and 0 otherwise ($\mathbf{r}_i(v)$ denotes the vector element 
corresponding to node $v$).
A restart parameter, $\alpha$ (also called damping factor) is used to tune the frequency at which
the walker ``teleports" back to the seed nodes.
The RWR-based proximity is defined as the steady state:
\begin{equation}
\label{rwr}
\mathbf{p}_i = (1-\alpha)\mathbf{A}^{(cs)}\mathbf{p}_i + \alpha\mathbf{r_i}
\end{equation}
where $\mathbf{p}_i(v)$ denotes the probability of being at node $v$ when the walk continues for
a sufficiently long time. 
The steady state vector $\mathbf{p}_i$ is used to rank nodes for prediction, where higher values $\mathbf{p}_i(v)$ correspond to higher likelihood that node $v$ is labeled the same as nodes of $S_i$. 
This procedure can be repeated for each label set $S_i, i=1...n$ and the most likely label 
(i.e. the $\mathbf{p}_i(v)$ with highest value) is predicted for node $v$.

\vspace{0.02in}
\noindent
{\bf Random walks with symmetric degree normalization:}
While the above formulation of RWR is intuitive, a different normalization technique is often used to 
scale the transition probabilities by the in- and out-degree of nodes~\cite{degnorm}.
This ``symmetric" normalization technique uses transition matrix $\mathbf{A}^{(sym)}$, 
where $\mathbf{A}^{(sym)}=\mathbf{D}^{-1/2}\mathbf{A}\mathbf{D}^{-1/2}$ and $\mathbf{D}_{i,i} = \sum_{k}{\mathbf{A}_{i,k}}$. 
Since $\mathbf{A}^{(sym)}$ is not a stochastic matrix, a re-normalization step is introduced to the 
RWR formulation to produce the probability vector $\mathbf{p}$:
\begin{equation}
\label{rwr2}
\begin{aligned}
\mathbf{\hat{p}} & = & (1-\alpha)\mathbf{A}^{(sym)}\mathbf{p} + \alpha\mathbf{r} \\
\mathbf{p} & = & \mathbf{\hat{p}} / |\mathbf{\hat{p}}|
\end{aligned}
\end{equation}

\vspace{0.02in}
\noindent
{\bf Label propagation with negatively-labeled examples:} 
In some applications, a set of negatively-labeled nodes $N_i$ (i.e., anti-labels that specify a node is {\em not} 
of a specific class) is provided. 
When such information is not available, it is also potentially useful to sample negatively-labeled nodes from nodes that are not positively labeled (e.g. selecting $N_i$ as a subset of $\cup_{i \neq j}S_i$) and use them to inform label propagation.
The objective of label propagation with negative examples is to predict labels for unlabeled nodes that do not contradict the anti-labels.
This is illustrated in Figure~\ref{fig:labelprop}.

Many existing methods for label propagation utilize negative examples by formulating 
an optimization problem where the objective function penalizes predicting positive labels for 
negatively labeled nodes~\cite{SRW, TELETUNE, RWER, QUINT}.
({\sc SRW}), one of the earliest algorithms that considers negative examples,  
learns a function to optimize edge weights such that positive examples are ranked 
higher than negative examples~\cite{SRW}. 
This is accomplished by embedding the restart vector $\mathbf{r}$ into the 
transition matrix $\mathbf{A}$ and explicitly restricting updates that would alter the matrix 
elements corresponding to $\mathbf{r}$. 
A more recent work on query-specific optimal networks ({\sc QUINT}) takes a similar 
approach to adjusting the weight -- or existence -- of edges defined by $\mathbf{A}$, 
but it formulates the problem in terms of a single positive example (i.e. $|S_i| = 1$) and 
ignores the restart probability as a scaling factor \cite{QUINT}. 
The teleportation tuning method of Berberidis {\em et al.} learns a weighted restart vector $\mathbf{r}_i$ for each label $S_i$ that optimizes within-class predictions \cite{TELETUNE}, but this results in a model where all nodes restart to a given node with the same probability. 
More recently, random walk with extended restarts ({\sc RWER}) attempts to learn an optimal restart probability for each node $v \in V$\cite{RWER} for a specific $S_i$. 
However, the method scales the strength of all edges incident to a node uniformly in relation to the restart probability, resulting in no discrepancy between positive and negative neighbors.

\subsection{Proposed Approach: Combining Edge Re-weighting and Restart Tuning.}
We propose to combine the ideas of edge re-weighting and restart tuning such that: 
(i) the walker restarts with higher probability ($>\alpha$) when it encounters an edge leading to a negatively labeled node, but
(ii) continues walking with the default probability ($1-\alpha$) when it encounters an edge leading to an unlabeled or positively labeled node. 
This has several benefits:
1) It does not artificially inflate the rank of nodes by redirecting the walker to a smaller group 
of neighbors.  
2) It does not reduce the rank of unlabeled neighbor nodes by avoiding them in an effort to avoid 
the negatively labeled node. 

Here, we develop a framework to realize this approach by reformulating RWR 
in an intuitive way that creates a single transition matrix composed of ``restart edges" 
and ``transition edges". 
We then adjust the entries of these matrices based on the given set of positive ($S_i$) and
negative ($N_i$) examples.

\begin{figure*}[t]
 \begin{center}
  \includegraphics[width=\textwidth]{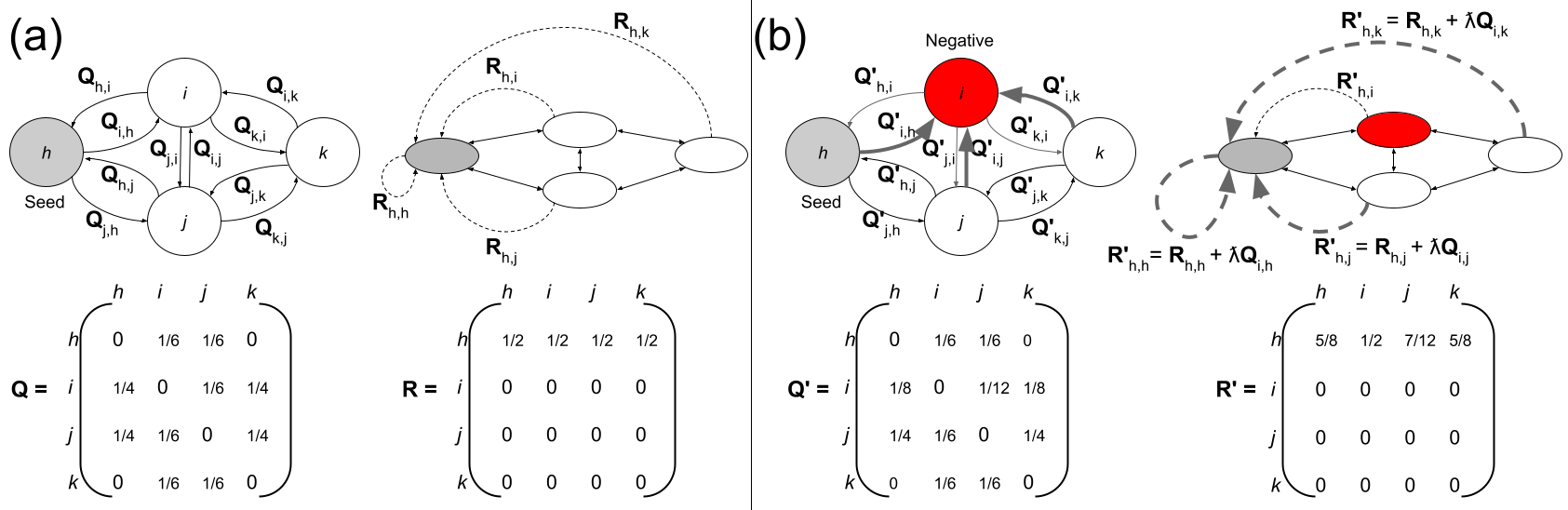}
 \end{center}
  \vspace{-0.1in}
  \caption{{\small{{\bf Reformulation of random walks using transition and teleport matrices to allow variable restarts}. The seed node is shown in gray. The transition edges are labeled by $\mathbf{Q}$, the teleport edges
  are labeled by $\mathbf{R}$. (a) A reformulated random walk with restarts that is equivalent to the classical formulation with $\alpha=0.5$ where columns of $\mathbf{Q}$ are column normalized. Note that the row $\mathbf{R}_{h,:}$ that corresponds to the seed node $h$ contains all uniform entries. (b) The random walk modified to avoid negative node $i$ using a \dampen~factor $\lambda=0.5$, where re-weighted edges have been highlighted in bold and the updated matrices $\mathbf{Q'}$ and $\mathbf{R'}$ are shown below. The edges that lead to the negative node have been re-weighted as $\mathbf{Q}'_{i,:} = (1-\lambda)\mathbf{Q}_{i,:}$. The restart edges leaving nodes $v \in Adj(i)$ have been updated as $\mathbf{R'}_{h,v} = \mathbf{R}_{h,v}+\lambda \mathbf{Q}_{i,v}$ to direct the walker back to the seed rather than transitioning to the negative node. This formulation allows restarting with different probabilities depending
  on the current node visited by the walk.}}}
  \label{qandr}
\end{figure*}

\vspace{0.02in}
\noindent
{\bf Reformulation of random walks to unify transition and teleport:}
Considering the classical RWR formulation, the first term on the right-hand-side of Equation~\ref{rwr} 
captures the transition of the random walker from the current node to adjacent nodes, and the second term captures 
the random walker ``teleporting'' to seed nodes. 
Observing that $|\mathbf{p}| = 1$ by definition, 
we can express $\alpha\mathbf{r}$ as:
\begin{equation*}
	\alpha\mathbf{r} = \alpha\mathbf{r}\mathbf{1}^{T}\mathbf{p}
\end{equation*}
\noindent
where $\mathbf{1}^T$ is a row vector of all 1's of compatible dimension to $\mathbf{r}$ such that $\alpha\mathbf{r}\mathbf{1}^T = \mathbf{R} \in \mathcal{R}^{|V| \times |V|}$. 
Noting that $\mathbf{Rp}=\alpha\mathbf{r}$ and setting 
$\mathbf{Q}^{(cs)}=(1-\alpha)\mathbf{A}^{(cs)}$, we can rearrange Equation~\ref{rwr} as an ordinary eigenvector equation:
\begin{equation}
\label{eve}
\mathbf{p} = (\mathbf{Q}^{(cs)} + \mathbf{R})\mathbf{p},
\end{equation}
\noindent
where $\mathbf{Q}^{(cs)}$ captures the transition of the random walker to adjacent nodes 
and $\mathbf{R}$ captures teleport to seed nodes. 
The intuition behind this formulation $\mathbf{R}$ is illustrated in Figure~\ref{qandr}, 
where the reformulation effectively adds an edge from every $v \in V$ to every $u \in S$ with 
transition probability $\alpha$.
Similarly, for random walks with symmetric normalization, Equation~\ref{rwr2} can be reformulated 
as:
\begin{equation}
\label{eve2}
\begin{aligned}
\mathbf{\hat{p}} & = & (\mathbf{Q}^{(sym)} + \mathbf{R})\mathbf{p} \\
\mathbf{p} & = & \mathbf{\hat{p}} / |\mathbf{\hat{p}}|
\end{aligned}
\end{equation}
where $\mathbf{Q}^{(sym)}=(1-\alpha)\mathbf{A}^{(sym)}$.
In our implementation, we use this reformulation of symmetric random walk.

\vspace{0.02in}
\noindent
{\bf Variable restarts:}
Consider a more flexible model where rather than the walker restarting with a fixed probability $\alpha$ at every node, 
the walker is free to restart with a unique probability depending on where the walker is at. 
This flexibility can be directly incorporated into the above formulation, since each entry of $\mathbf{R}$ 
represents a directed edge from a given node to a seed node.
The immediate benefit to such a model is it allows the walker to restart to a seed whenever it encounters an edge leading to a negative example.
i.e,
given a node $u \in N_i$, the values $\mathbf{Q}_{u,v}$ for all $v \in Adj(u)$ can be reduced, 
and the difference distributed among the restart edges $\mathbf{R}_{w,v}$ for $w \in S_i$.  

\vspace{0.02in}
\noindent
{\bf Adjusting restart and transition edges based on negative examples:}
Let  $u \in N_i$ be a negatively-labeled example for label $i$.
For each $v$ adjacent to $u$, we reduce transition probability from $v$ to $u$ and redistribute these probabilities to the seed vertices $S_i$ as
follows:
\begin{equation}
\begin{aligned}
    \mathbf{R}_{s,v} = \mathbf{R}_{s,v}+\frac{\lambda\mathbf{Q}_{u,v}}{|S_i|}~\text{if~$v \in Adj(u)$~and~$s\in S_i$} \\
    \mathbf{Q}_{u,v} = (1-\lambda) \mathbf{Q}_{u,v}~\text{if~$v \in Adj(u)$}
\end{aligned}
\label{eq:adjust}
\end{equation}
where $\lambda$ is a ``\dampen'' parameter used to tune the degree of aggressiveness in steering the walk away from 
negatively-labeled nodes.  
In the next section, we comprehensively characterize the effect of $\lambda$ on predictive accuracy.
Observe that this adjustment retains the sum of the $v$th column of $\mathbf{Q}+\mathbf{R}$. 

\subsection{Label Propagation via \algoname.}
The matrix $\mathbf{R}$ is independent of the label that is to be propagated, thus we first construct 
$\mathbf{R}$ based on the input graph $G(V,E)$.
Then, for each label $i$ with set $S_i$ of positively-labeled nodes, we first construct the matrix 
$\mathbf{Q}^{(sym)}$.
If negatively-labeled nodes are not available, we sample negatively-labeled nodes from $V\setminus S_i$
to obtain $N_i$, using the methodology described in the next subsection.
Subsequently, we adjust  $\mathbf{R}$ and $\mathbf{Q}^{(sym)}$ based on $N_i$, using Equation~\ref{eq:adjust}.
We then compute $\mathbf{p}_i$ using Equation~\ref{eve2} and rank the nodes in $V\setminus S_i$ according to 
this vector to prioritize the assignment of label $i$.

\subsection{Sampling Negatively Labeled Nodes.}
\label{negsampsec}
If a set of negatively-labeled nodes is not available, it is necessary to sample negatively-labeled nodes 
from the set of nodes that are not positively-labeled.
In the literature, negative sampling methods have been proposed based on prioritizing confident false predictions~\cite{PNLP}. 
It follows that false negatives are nodes those that are close to one or more seed nodes. 
For this reason, it can be good strategy to select negatively-labeled examples from the set of nodes that are in the neighborhood
of positively-labeled nodes.
To investigate how proximity of the sampled negatively-labeled nodes to seeds affects predictive performance,
we sample negatives from the nodes uniquely reachable in exactly $k$-hops from each seed node. 
For this purpose, to generate a pool of candidate negatively-labeled nodes, we use breadth-first search 
and identify nodes that (i) are at depth of $k$ hops from the seeds, and (ii) do not have the same
label as the seed. 
From this pool, we draw uniformly at random a sample that is of size at most as (if possible, equal to) 
the number of seeds (positively-labeled nodes).
This ensures that the sets of positively and negatively labeled nodes are as balanced as possible.


\section{Results}
\label{sec:results}

\subsection{Experimental Setup.}

We evaluate the predictive performance of \algoname\ against existing methods using 
multiple network datasets that are often used to benchmark label propagation and node classification
algorithms.
These datasets include the CORA dataset~\cite{CORA}, a CiteSeer dataset~\cite{CITESEER}, the Political Blogosphere 
dataset~\cite{POLBLOGS} and a Facebook dataset~\cite{FB1}.
The characteristics are summarized in Table \ref{datasets}. 
For consistency, we convert networks with directed edges to undirected networks, and remove 
nodes that are isolated from the rest of the network.

\begin{table}[t]
\caption{{\small{\bf Network datasets with node labels used to evaluate label prediction performance.}}}
    \begin{center}
     \begin{tabular}{||c | c | c | c||} 
     \hline
     Name & \# Nodes & \# Edges & \# Labels \\ [0.5ex] 
     \hline\hline
     CiteSeer & 3312 & 4660 & 6 \\
     \hline
     CORA & 2,708 & 10,556 & 7 \\
     \hline
     Polblogs & 1,224 & 16,718 & 2 \\
     \hline
     Facebook & 22,470 & 171,002 & 4 \\
     \hline
    \end{tabular}
    
    \label{datasets}
    \end{center}
\end{table}

\vspace{0.02in}
\noindent
{\bf Sampling of training and validation sets:}
In our experiments, we consider the case where training data is scarce, i.e., most of the labels in the network are unknown.
Namely, from each set of labeled nodes $S_i \in \mathcal{S}$ for a given network, we sample, uniformly at random, 50 
positive training (seed) sets $s_{1},s_{2},...,s_{50}$ of fraction $\gamma$ of the nodes in $S_i$, e.g. $s_j \subset S_i$ and $|s_j| = \gamma*|S_i|$. 
For each seed set $s_{j}$, we draw up-to the same number of negative training sets $n_{j}$ at distances $k=[1,2,3]$ from the seeds
using the strategies outlined in Section~\ref{negsampsec}. 
Due to network topology and the location of the nodes in $s_j$, there are cases where $|n_j|<|s_j|$, we perform the experiment 
as long as $|n_j|>0$. 
If $|n_j|=0$, we sample a new seed set $s_j$ until at least one negatively labeled node at distance $k$ 
can be found. 
We use the set $T_j = \{s_j \cup~n_j\}$ for training, leaving  $V \backslash T_j$ for validation.

\vspace{0.02in}
\noindent
{\bf Parameter settings:}
We determine through an initial parameter sweep that the RWR based methods perform optimally with 
a restart probability $\alpha=0.05$, thus we use this value in all experiments. 
During our baseline accuracy assessment, we set \algoname's \dampen~parameter to $\lambda=0.9$ based on initial experiments that showed higher values of this parameter provide better predictive performance. 
We perform two additional experiments to characterize the effects of the \dampen~factor $\lambda$ and the training set size $\gamma$. We varied $\lambda$ over the values $[0.2, 0.4, 0.6, 0.8, 1.0]$ and $\gamma$ over the values $[0.02, 0.05, 1.0]$. 

\vspace{0.02in}
\noindent
{\bf Competing methods:}
We compare the predictive performance of \algoname\ against classical RWR with symmetric normalization~\cite{RWR}, 
{\sc QUINT}~\cite{QUINT}, and {\sc RWER}~\cite{RWER}, where the latter two methods learn optimal transition strategies 
using gradient descent. 
For {\sc QUINT}, the authors provide several variations and we select their first order Taylor polynomial 
approximation as all three variations show equivalent performance in the benchmark experiments reported by 
the authors~\cite{QUINT}.

In \algoname, the positively labeled training nodes and the seed set are identical.
This is not the case for {\sc QUINT} and {\sc RWER}. 
For both algorithms, the setting involves sets of positive and negative example nodes, as well as
a single query (seed) node (i.e. $|s_i|=1$). 
The methods then learn optimal networks or restart profiles that rank the positive nodes higher than the negative nodes 
while propagating the label only from the single query node. 
This makes direct comparison to our set-based method problematic, so we create a modified version of our 
method that also works with a single query node. 
The modified \algoname$_{sq}$ accepts the same inputs as {\sc QUINT} and {\sc RWER}, but adds 
edges to $G$ between the query node and the positively-labeled training nodes 
before applying the edge-weight redistribution for negatively-labeled training nodes. 
This allows us to propagate the label from a single query node, but leverage the positive nodes in a way that 
is similar to treating them as additional seed nodes.

\begin{figure*}[t]
    \centering
    \includegraphics[width=0.7\textwidth]{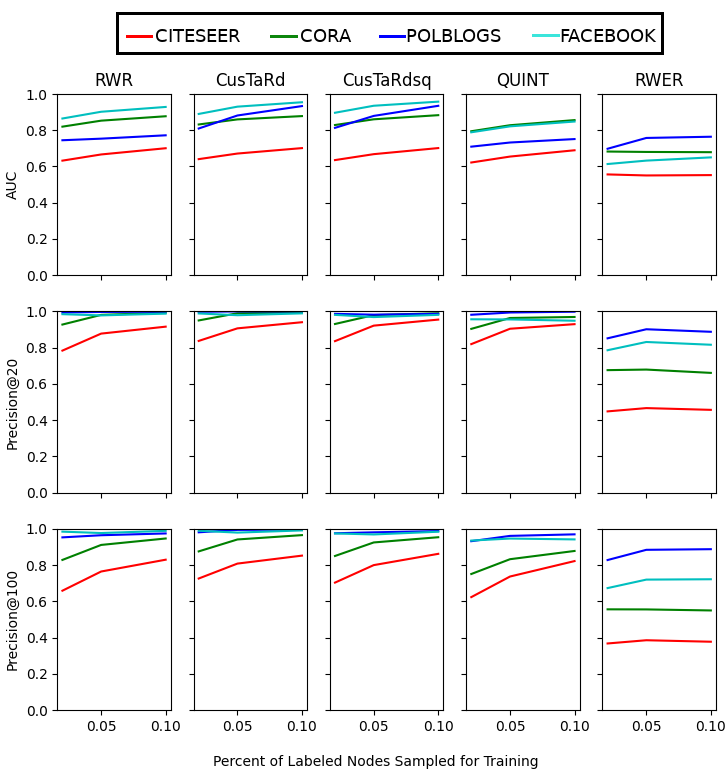}    
    \caption{{\small{{\bf Predictive performance of label propagation algorithms as a
    function of training set size.}
    Positive training sets are sampled of sizes 2\%, 5\% and 10\% of available positive examples
    (seed nodes) for each label. Negative examples are sampled to be of equal size from nodes at a $k$-hop distance of 1 to positive examples. The reported values are averages across 50 validation instances.}}}
    \label{fig:trainsize}
\end{figure*}

\vspace{0.02in}
\noindent
{\bf Evaluation of predictive performance.}
We use each method to propagate the labels of sample $s_j$ and then rank the nodes by confidence of predictions.
The node rankings are then evaluated from most confident to least confident, assigning ``true positive'' or ``false positive'' 
to each prediction. 
The Area Under ROC Curve (AUCs), Precision@20, and Precision@100 are computed by 
combining the TP/FP counts at each rank position for all $s_j$ across all labeled sets 
$S_i$ to generalize the performance for a given dataset.
We report the mean and standard deviation of these values across the 50 validation instances.

\subsection{Predictive Performance.}
The predictive performance of all algorithms on all four datasets are shown in
Figure~\ref{fig:trainsize} as a function of training set size, using three different performance criteria.
The average and standard deviation of the performance metrics for training size 2\% are also shown in in Table~\ref{table:perc2}. 
\algoname\ consistently achieves highest scores for Precision@20 and Precision@100, 
and for AUC the best performance is achieved by either \algoname\ or \algoname$_{sq}$. 

We observe that the  CiteSeer network is the most difficult dataset for all methods to deliver accurate predictions.
For this network, Precision@20 is in the low 80 percent range even for the best-performing algorithms. 
The minimum variance in prediction accuracy is displayed by \algoname\ for most datasets and metrics, 
with the exception of the CiteSeer network where it is higher or equal than one or more of the other methods.

\begin{table*}
\caption{{\small{{\bf Predictive performance of \algoname\ and competing methods on four benchmarking
datasets according to three different performance criteria}. 
For each algorithm, dataset, and performance metric, the mean performance metrics ± standard deviation 
is shown across $50$ randomly generated validation instances, with 2\% of positively-labeled nodes
selected for training, and negatives sampled at $k$-hop distance $1$.}}}
    \begin{center}
     \begin{tabular}{||c c c c c c||} 
     \hline
     Network & $RWR$ & \algoname & \algoname$_{sq}$ & $QUINT$ & $RWER$ \\ [0.5ex]
\hline\hline
\multicolumn{6}{||c||}{AUC} \\
 CiteSeer & 0.632±0.129 & {\bf 0.641±0.130} & 0.635±0.131 & 0.622±0.130 & 0.556±0.122 \\
 Cora     & 0.820±0.089 & {\bf 0.832±0.084} & 0.829±0.084 & 0.794±0.093 & 0.682±0.188 \\
 Polblogs & 0.745±0.051 & 0.810±0.050 & {\bf 0.813±0.057} & 0.709±0.049 & 0.698±0.190 \\
 Facebook & 0.865±0.037 & 0.890±0.034 & {\bf 0.897±0.035} & 0.789±0.063 & 0.613±0.142 \\
\multicolumn{6}{||c||}{Precision@20} \\
 CiteSeer & 0.784±0.236 & {\bf 0.837±0.243} & 0.836±0.247 & 0.820±0.250 & 0.448±0.502 \\
 Cora     & 0.927±0.112 & {\bf 0.950±0.093} & 0.931±0.123 & 0.904±0.120 & 0.676±0.398 \\
 Polblogs & 0.994±0.026 & {\bf 0.998±0.012} & 0.987±0.043 & 0.982±0.043 & 0.852±0.320 \\
 Facebook & 0.985±0.039 & {\bf 0.989±0.026} & 0.981±0.038 & 0.957±0.083 & 0.786±0.353 \\
\multicolumn{6}{||c||}{Precision@100} \\
 CiteSeer & 0.658±0.250 & {\bf 0.726±0.272} & 0.703±0.266 & 0.623±0.252 & 0.367±0.392 \\
 Cora     & 0.828±0.145 & {\bf 0.875±0.133} & 0.850±0.146 & 0.751±0.149 & 0.556±0.363 \\
 Polblogs & 0.953±0.029 & {\bf 0.981±0.018} & 0.976±0.029 & 0.932±0.051 & 0.828±0.306 \\
 Facebook & 0.984±0.020 & {\bf 0.989±0.013} & 0.975±0.033 & 0.935±0.082 & 0.673±0.401 \\
     \hline
    \end{tabular}
    
    \label{table:perc2}
    \end{center}
\end{table*}

Comparison of the left-most column of Table~\ref{table:perc2} against other columns of the table
shows that algorithms that utilize negative samples deliver more accurate predictions (with the exception of RWER).
This implies that methods based on post-filtering negatives from results will yield less accurate predictions 
than methods which incorporate the negatives into the ranking procedure. 
This may be particularly useful in applications where a large number of negatives are available, such as  
differential expression studies in biology where many genes or proteins have no significance.

\subsection{Effect of Sampling of Negative Examples.}
Figure \ref{fig:kandlambda} (left) plots the different performance metrics versus the $k$-hop proximity of negative examples for all four networks. It shows for all methods except RWER that negatives at $k$-hop proximity 1 to the seeds result in optimal performance. For RWR, $\algoname$, $\algoname_{sq}$ and some QUINT results, performance was inversely correlated with $k$ (i.e. performance decreased as $k$-hop distance increased). However, some QUINT results exhibited lowest performance at $k$-hop distance 2 rather than 3, making them less correlated but still consistent with the observation that the negatives sample at distance 1 are most informative. RWER achieved optimal performance at $k$-hop distance 2, though the performance was still lower than the optimal performance of $\algoname$ and $\algoname_{sq}$. Based on the results, it would be reasonable to sample negatives as close to the seeds as possible. This behavior has the nice property of limiting the neighborhood of nodes that must be evaluated in the search to manually annotate negatives.

\subsection{Effect of Redirection Factor.}
\begin{figure*}[t]
    \centering
    \includegraphics[width=\textwidth]{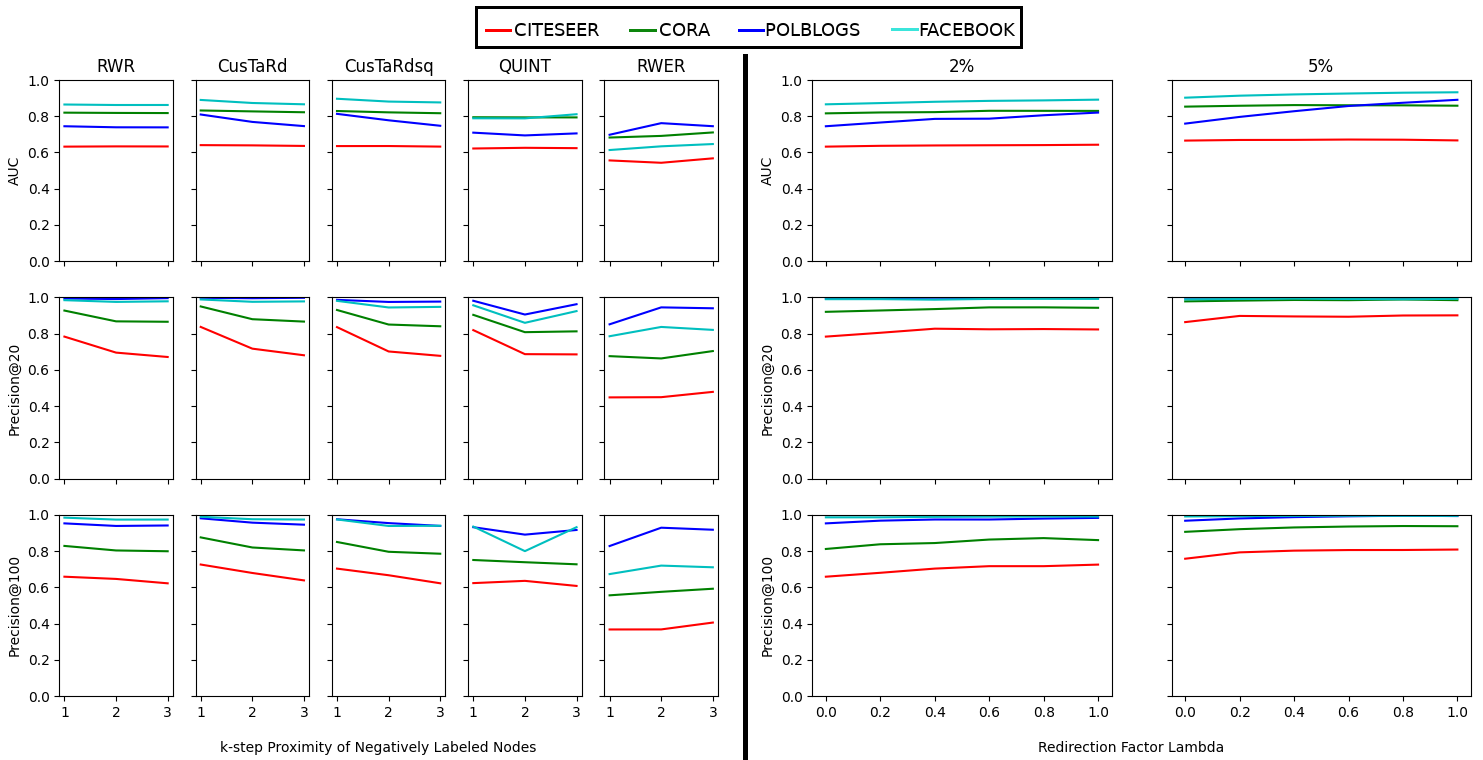}    
    \caption{{\small{{\bf The effects of sampling of negative examples and redirecting edge weights on predictive performance}
    {\bf Left: }As discussed in Section~\ref{negsampsec}, we sample negatively-labeled training nodes from
    the set of nodes that are not positively labeled, by constructing pools of candidate nodes
    based on their distance to positively-labeled nodes. The curves show the effect of this 
    distance on predictive performance for all algorithms on all datasets according to all performance
    criteria that are considered.
    {\bf Right:} Our reformulated random walk depends on the \dampen~factor $\lambda$ as defined in Equation \ref{eq:adjust}. The plot shows the effects of varying the \dampen~factor for different sizes of training data. The value of $\lambda$ was varied over $[0.0, 0.2, 0.4, 0.6, 0.8, 1.0]$. Training sets were sampled of sizes 2\% and 5\% for each label and negative nodes were sampled at a $k$-hop distance of 1.
    }}}
    \label{fig:kandlambda}
\end{figure*}

Figure \ref{fig:kandlambda} (right) plots the performance metrics for $\algoname$ versus the \dampen~factor $\lambda$ for different sample sizes and fixed negative node $k$-hop distance 1. The overall response curve shapes do not exhibit significant differences when varying the sample size, but the curves are quite different between networks. The AUC curves for CORA and CITESEER show slight decreases in performance at the highest values of $\lambda$, while the POLBLOGS result shows increasing performance all the way to $\lambda=1.0$. The gain in performance is more pronounced for Precision@100 than for Precision@20 showing that increasing $\lambda$ helps to increase the ranking of more distant nodes.

\section{Conclusion}
\label{sec:discussion}
In this study, we reformulated random walks to enable variable restarts, 
which in turn gave rise to \algoname, an algorithm for effectively utilizing
negatively-labeled nodes in label propagation.
\algoname\ does not ``learn" parameters or solve an optimization problem, it uses a single
parameter to directly modify the entries of the stochastic matrix to redirect flow from negatively-labeled
nodes to positively-label nodes.
In addition to reformulation of random walks, \algoname\ samples negatively-labeled
nodes from the neighborhood of positively-labeled nodes, thereby learning to discriminate
between positively and negatively labeled nodes.
Our comprehensive experiments on four benchmark networks showed that \algoname\ consistently
outperforms competing optimization/learning-based algorithms, and its predictions are 
consistently robust to scarce training samples.
Finally, our systematic experiments showed that sampling negative examples in the neighborhood
of positive examples improves prediction accuracy for all algorithms.

These results lay the foundations for more effective incorporation of 
label propagation into machine learning frameworks.
Integration of the algorithm described here with machine learning models
that use node features can further improve the accuracy and robustness
of such models.

\bibliographystyle{plain}
\bibliography{bibliography.bib}

\end{document}